\documentclass[preprint]{aastex}

\usepackage{natbib}
\usepackage{amsmath} \usepackage{amssymb}
\usepackage[colorlinks=true,urlcolor=black,linkcolor=black,citecolor=black]{hyperref}
\usepackage{ulem}

\shorttitle{Time Delay Analysis of the Lensed Quasar SDSS J1001+5027}
\shortauthors{Aghamousa, Shafieloo}

 \usepackage{xcolor}
 \newcounter{attnctr} 
 
 \newcounter{delctr} 
 
 \newcounter{addctr} 

\begin{document}

\title{Time Delay Analysis of the Lensed Quasar SDSS J1001+5027}

\author{Amir Aghamousa}
\affil{Korea Astronomy and Space Science Institute, Daejeon, 305-348 Korea}
\affil{Asia Pacific Center for Theoretical Physics, Pohang, Gyeongbuk 790-784, Korea}
\email{amir@aghamousa.com}

\author{Arman Shafieloo}
\affil{Korea Astronomy and Space Science Institute, Daejeon, 305-348 Korea}
\affil{University of Science and Technology, Daejeon 34113, Korea}
\email{shafieloo@kasi.re.kr}

\begin{abstract}
We modify the algorithm we proposed before in \cite{aghamousa_timedelay_1} on time delay estimation of the strong lens systems incorporating weighted cross-correlation and weighted summation of correlation coefficients. We show the high performance of this algorithm by applying it on Time Delay Challenge (TDC1) simulated data.
We apply then our proposed method on the light curves of the lensed quasar SDSS J1001+5027 since this system has been well studied by other groups to compare our results with their findings. 
In this work we propose a new estimator namely ``mirror" estimator along with a list of criteria for reliability test of estimation.
Our mirror estimator results to $-117.1^{+7.1}_{-3.7}$ and $-117.1^{+7.2}_{-8.8}$ using simple Monte Carlo simulations and simulated light curves provided by \cite{COSMOGRAIL_XIV} respectively. 
Although TDC1 simulations do not reflect the properties of SDSS J1001+5027 light curves, using these simulations results to smaller uncertainty which shows the higher quality
observations can lead to substantially more precise time delay estimation.
Our time delay estimation is in agreement with findings of the other groups for this strong lens system and the difference in the size of the error bars reflects the importance of appropriate light curve simulations. 
\end{abstract}

\keywords{}

\section{Introduction}\label{introduction}

Cosmology has moved towards becoming a precision science in the last few decades where cosmologists have now access to variety of astronomical data.
Strong gravitational lensing systems contain a wealth of information that can be used to estimate the expansion history of the universe \citep{Eric2004, Suyu2013}. In a strong gravitational lensing system the light rays of a source quasar propagate toward us experiencing the gravitation of a lensing galaxy along the different optical paths which result in multiple images. The light curves associated with these images show time lags due to differences in optical paths and the gravitational potential of the lensing galaxy \citep{Refsdal1964, Kochanek2002, Treu_review}. These information can be used to measure cosmological distances through so-called time delay estimation.
An accurate time delay estimation in combination with a proper modeling of the lens galaxy can lead to an independent measurement of the Hubble parameter \citep{Suyu2009, Suyu2013}. Using strong lens systems we can reconstruct the expansion history of the universe in a complementary approach in comparison to the way we measure the expansion history using standardized candles and rulers (using supernova data or baryon acoustic oscillation) \citep{Eric2004, Eric2011, Suyu2012}. To obtain this goal, we need precise and accurate estimation of the time-delays \citep{Tewes2013a, Tewes2013b, Hojjati2014, aghamousa_timedelay_1} as well as a realistic modeling of the lensing systems \citep{Oguri2007, Suyu2009, Suyu2010, Fadely2010}. 

In this paper we present an improved version of our previous time delay estimation algorithm that was proposed in \cite{aghamousa_timedelay_1} and was shown to be reliable in the Time Delay Challenge (TDC) \citep{TDC1-2014}, applying it on a real data sample. In this new algorithm we enhance the accuracy and precision of the estimator by using the error-sensitive weighted cross-correlation in the analysis (rather than simple cross-correlation) and implementing a new procedure for combining correlations with some additional statistical criteria. Using these additional tools we offer a new time delay estimator along with a procedure of error estimation. The results of applying new algorithm on TDC1 simulations show dramatically improvements in accuracy and completeness metrics in comparison with previous version.

We apply this algorithm on doubly lensed quasar SDSS J1001+5027 \citep{Oguri2005} which is one of strong lens systems monitored in a long term by COSMOGRAIL collaboration (http://www.cosmograil.org) and results in two light curves of images which are publicly available \footnote{http://cosmograil.epfl.ch/data} \citep{COSMOGRAIL_XIV}. The COSMOGRAIL project has gathered light curves of tens of lensed quasars and proposed several techniques for estimating the associated time delays. For this particular system (SDSS J1001+5027) they have presented time delay measurements using five different techniques including three techniques introduced by \cite{Tewes2013a}, one by \cite{Hojjati2013} and a new algorithm by the collaboration authors. Considering the fact that the lensed quasar SDSS J1001+5027 is the most well studied system in the literature we decided to focus analyzing this system to compare our results with other existing ones.   

In the following we first describe the light curves data of SDSS J1001+5027 system (Section~\ref{sec:data}). Next in Section~\ref{sec:methodology} we start elaborating the methodology of our analysis which contains smoothing method (Section~\ref{subsec:smoothing}), weighted cross-correlation (Section~\ref{subsec:cross}), explanation of time delay estimation procedure (Section~\ref{subsec:time_delay}) and the  evaluation of performance (Section~\ref{subsec:performance}).
This section continues with an error estimation procedure based on distribution of estimations (Section~\ref{subsec:error_estimation}) and explanations of used simulations (Section~\ref{subsec:simulations}).
We present and discuss the results of the analysis in Section~\ref{sec:results} and finally we conclude in Section~\ref{sec:conclusion}.

\section{Data}~\label{sec:data}
\begin{figure}
   \includegraphics[width=\textwidth]{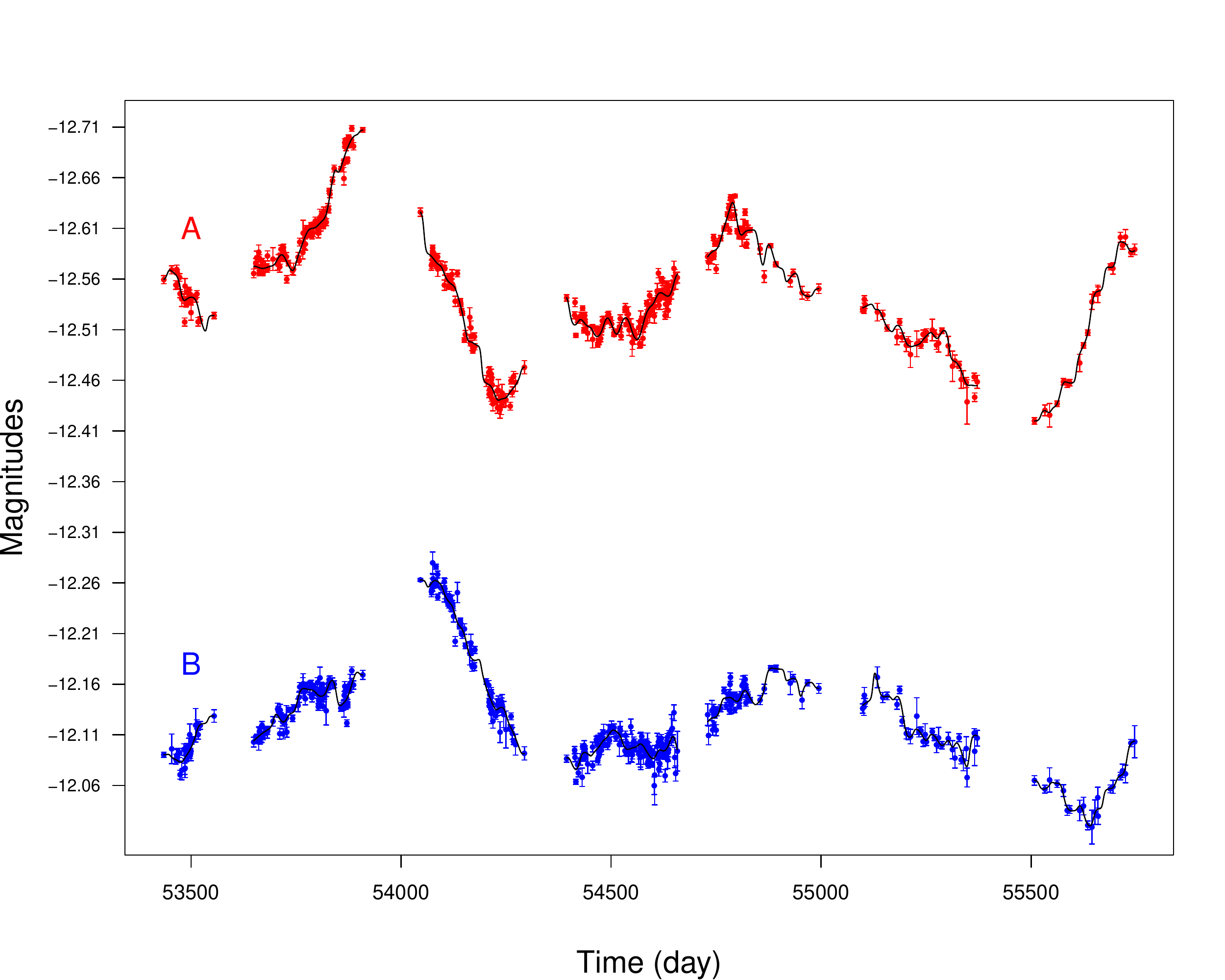}
  \caption{\label{fig:data1} The light curves of lensed quasar SDSS J1001+5027 provided by \cite{COSMOGRAIL_XIV}. Due to seasonal gaps the light curves appear in segments of data. The solid curves are the corresponding smooth fits using smoothing method explained in Section~\ref{subsec:smoothing} with $\Delta=8$ days. We ignore the smooth fit in gap regions due to lack of data.}
 \end{figure}

The SDSS J1001+5027 is a doubly lensed quasar at z=1.838 ($\alpha_{2000}=10:01:28.61, \delta_{2000}=+50:27:56.90$) \citep{Oguri2005}. 
The COSMOGRAIL collaboration monitored this system for nearly six years through three different telescopes \citep{COSMOGRAIL_XIV} and finally they combined and corrected the light curves from these observations providing with a pair of light curves for 443 independent epochs. 
Figure~\ref{fig:data1} illustrates the light curves A (red points) and B (blue points). The time epochs of light curves are not equispaced and also have six gaps with the length more than 70 days. These gaps divide the light curves in seven different \textit{segments of data}. In addition each data light curve has different error bar in different epoch of time (this relates to an improvement in our algorithm in this work). In the next Section we will elaborate on our improved time delay estimation algorithm which considers all of these characteristics of the data.

\section{Methodology}~\label{sec:methodology}
In principle the time delay between two light curves of a strong lens system can be estimated by simply comparing them with each other looking at particular features in the data that are shifted in time with respect to each other. However observations have generally some limitations due to different practical issues so it is not so trivial to find such features in the light curves and measure the time delay in a straightforward way. Hence algorithms should be designed in a way to consider important characteristics of the data to get a reliable estimation. In this work we adopt and modify the algorithm proposed in \cite{aghamousa_timedelay_1} for our analysis which consists of a smoothing method and cross-correlation to find the time lag between light curves.
Here we use the similar smoothing method to obtain the smooth light curves versus a regular grid of time then we make some important modifications in the method in computing the cross-correlation, combing the correlation values and also in error estimation. We employ weighted cross-correlation that is error-sensitive and this is an important characteristic since the actual data has different uncertainties for different data epoch. We use TDC1 double systems simulated data to evaluate the performance of the algorithm. We also employ different extensive sets of simulations to derive the uncertainty profile of the time delay. 

\subsection{Smoothing method}~\label{subsec:smoothing}
We choose the same smoothing methodology used in \cite{aghamousa_timedelay_1} that was proposed by \cite{Arman2006,Arman2007,Arman2010} and \cite{Arman2012} to obtain smooth light curves at a regular fine grid of time. This method estimates the smooth light curve $A^{s}(t)$ at the time $t$ by an iterative algorithm:

\begin{equation}\label{equ:smooth}
A^{s}(t)= A^{g}(t) + N(t)\sum_{i}\frac{A^{d}(t_i)-A^{g}(t_i)}{\sigma^{2}_{d}(t_i)} \times exp\left [ - \frac{(t_i - t)^2}{2\Delta^2} \right ]
\end{equation}
where
\begin{equation}\label{equ:N}
N(t)^{-1}= \sum_{i} exp \left [ - \frac{(t_i - t)^2}{2\Delta^2} \right ] \frac{1}{\sigma^{2}_{d}(t_i)}
\end{equation}

The procedure starts with an initial guess model $A^{g}(t)=A^{g0}(t)$ and then the first smooth curve $A^{s1}(t)=A^{s}(t)$ will be obtained through Equation~\ref{equ:smooth} where $A^{d}(t_i)$ and $\sigma_{d}(t_i)$ are the $i^{th}$ data point and uncertainty in light curve data and $\Delta$ is smoothing width. $N(t)$ is the normalization factor which is calculated via Equation~\ref{equ:N}. The new smooth fit $A^{s2}(t)$ is obtained in the next iteration by substituting $A^{g}(t)=A^{s1}(t)$ in Equation~\ref{equ:smooth}. The procedure can be continued to higher iterations. While the final result is almost independent of the initial guess model $A^{g0}(t)$ \citep{Arman2007}, the smoothing width, $\Delta$, plays important role in controlling the smoothness of the fit. In other words, the smoothing width determines the smallest size of the fine structure in the data which appears in the final smooth fit. We should note that the smooth fit has dependency on the number of iterations $n_i$ as well but there is a mild degeneracy between width of smoothing and $n_i$ so by fixing one of them (preferably $n_i$) we can calibrate the other one to find the optimum combination \citep{Arman2007}. In this work we fix $n_i=3$ and $\Delta=8$ similar to values used in \cite{aghamousa_timedelay_1} and treat them as the fixed parameters in the analysis.

\subsection{Weighted cross-correlation}~\label{subsec:cross}

Using cross-correlation we can measure correlation coefficient between the two time series (light curves) corresponding to different time lags. 
The best time lag value can be obtained corresponding to maximum correlation coefficient \citep{Peterson2001}. Traditionally cross-correlation uses Pearson correlation coefficient, $\rho$, which reflects the linear dependency by a value between $-1$ to $+1$ \citep{Wasserman2004}. In practice the \textit{sample correlation coefficient, $r_{xy}$} is employed to estimate the $\rho_{xy}$,
\begin{equation}\label{equ:sample_cross}
	r_{xy}=\frac{\sum\limits_{i=1}^n (x_i-\bar{x})(y_i-\bar{y})}
            {\sqrt{\sum\limits_{i=1}^n (x_i-\bar{x})^2 \sum\limits_{i=1}^n (y_i-\bar{y})^2}},
\end{equation}
where $x_i, y_i, (i=1,2,..,n)$ are data samples (from two light curves in our case of study) with sample means $\bar{x}$ and $\bar{y}$ \citep{Peterson2001}.
In above equation all of the sample data points have the equal contribution in correlation coefficient while the importance of the data epochs could differ depending on their uncertainties. To improve our analysis, we utilize the \textit{sample weighted correlation coefficient} which estimates the correlation coefficient with assigning the weights $w_i$ to the sample data elements. The sample weighted correlation coefficient is given by
\begin{equation}\label{equ:sample_w_cross}
	r^{w}_{xy}=\frac{\sum\limits_{i=1}^n w_i (x_i-\bar{x}_w)(y_i-\bar{y}_w)}
            {\sqrt{\sum\limits_{i=1}^n w_i (x_i-\bar{x}_w)^2 \sum\limits_{i=1}^n w_i (y_i-\bar{y}_w)^2}},
\end{equation}
where $\bar{x}_w= \sum\limits_{i=1}^n w_i x_i$ and $\bar{y}_w= \sum\limits_{i=1}^n w_i y_i$ are the sample weighted mean of two data samples $x_i, y_i, (i=1,2,..,n)$ respectively. The weights $w_i$ are attributed to data epochs such that they can be normalized to one, $\sum\limits_{i=1}^n w_i=1$ \citep{wcorr}.
In Section~\ref{subsec:time_delay} we show how we include the error bar of the data in time delay analysis by using Equation~\ref{equ:sample_w_cross} with normalized weight equivalent to inverse of the variance of the data.

In implementing cross-correlation in our analysis we should note an important issue. As we discussed previously in \cite{aghamousa_timedelay_1}, correlation coefficient is invariant under linear transformation of variables \citep{Wasserman2004} and this property helps to tackle possible microlensing effects on data segments which we discuss in more details in Section~\ref{subsec:time_delay}. 
Another important issue in using cross-correlation in our case of study is about the number of data points involved in calculations (overlapped region between the two light curves). Obviously estimated correlation coefficients coming from less number of data epochs result to lower credibility. We will consider this fact in order to increase the reliability of the time delay estimation in our analysis.
\subsection{Time delay estimation}\label{subsec:time_delay}

Our algorithm for time delay estimation consists of using several statistical tools and criteria that can be described in following steps.

\begin{figure}
   \includegraphics[width=0.8\textwidth]{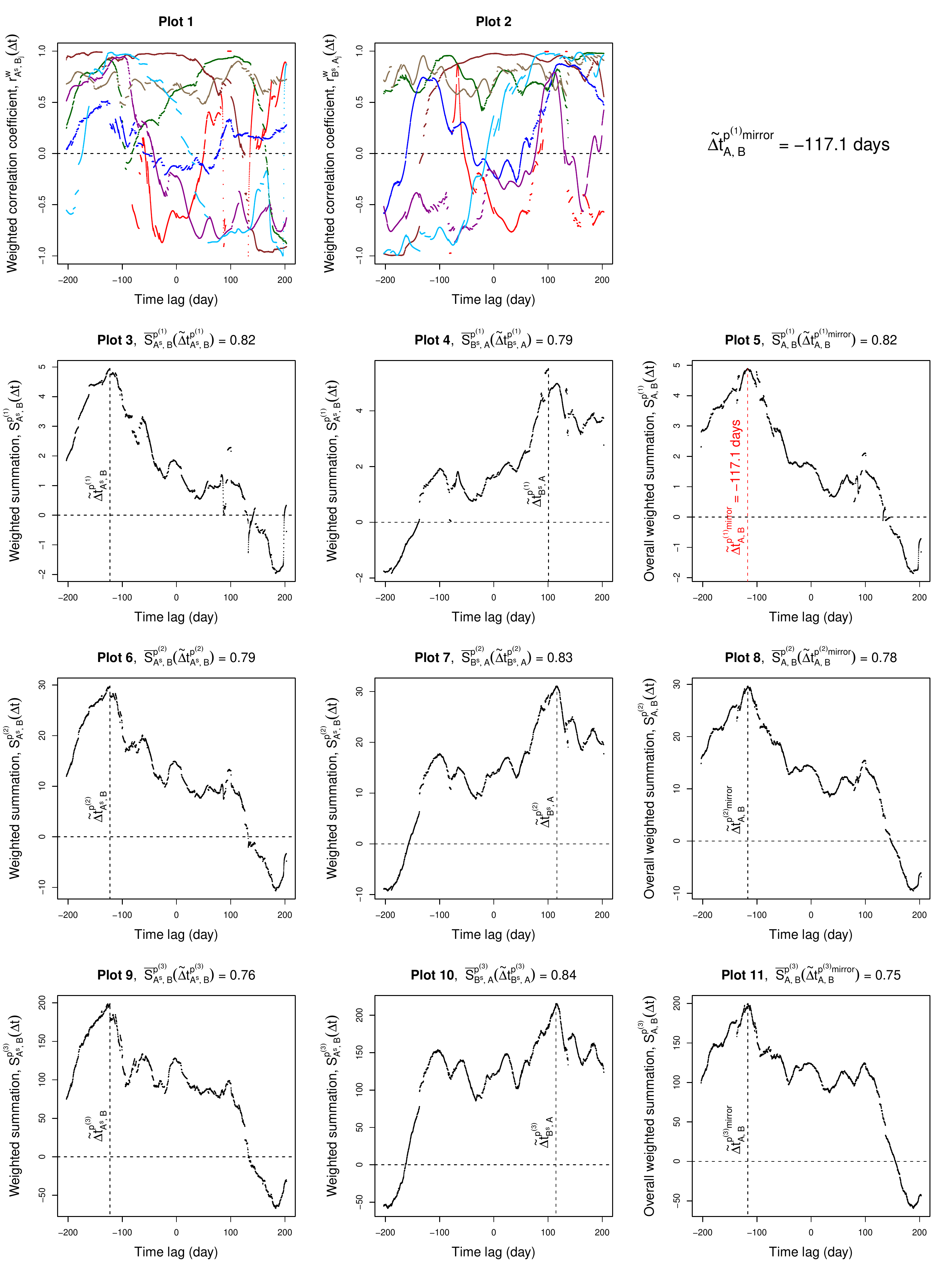}
  \caption{\label{fig:timedelay1} \footnotesize{Plot 1 and 2 show the weighted correlation coefficients $r_{A^s, B_j}^{w} (\Delta t)$ and $r_{B^s, A_j}^{w} (\Delta t)$ for seven data segments and smooth fit segments for the lensed quasar SDSS J1001+5027 light curves data respectively.
Plots 3, 6 and 9 represent the weighted summations i.e. $S_{A^s, B}^{p^{(1)}}(\Delta t), S_{A^s, B}^{p^{(2)}}(\Delta t), S_{A^s, B}^{p^{(3)}}(\Delta t)$ which are calculated from the weighted correlation coefficients shown in plot 1 using three different weights.
Similarly plots 4, 7 and 10 show $S_{B^s, A}^{p^{(1)}}(\Delta t), S_{B^s, A}^{p^{(2)}}(\Delta t), S_{B^s, A}^{p^{(3)}}(\Delta t)$ receptively calculated from the weighted correlation coefficients shown in plot 2.
Plot 5, 8 and 11 also represent the mirror time delay estimations associated to the pairs of weighted summation plots in the same row of plots.
The associated best time delay estimations are plotted with dashed vertical red lines while the corresponding weighted average correlations are written in the titles.
The mirror time delay estimation shown in plot 5, $\tilde{\Delta t}^{p^{(1)}\text{mirror}}_{A,B}=-117.1$, presents the time delay estimation in our algorithm considering passing all reliability criteria successfully.}
}
 \end{figure}

\begin{itemize}
  \item
	\textbf{Smoothing light curves, $A^s, B^s$:}
We utilize the smoothing method introduced in Section~\ref{subsec:smoothing} to obtain the smooth fits $A^s, B^s$ corresponding to light curves $A$ and $B$ respectively (Figure~\ref{fig:data1}). Since there are no data in the gaps between the data segments we ignore the smooth fit in gap regions and only consider the smooth fits corresponding to data segments.
Therefore as the light curves $A, B$ consist of data segments $A_j, B_j$, the associated smooth fits $A^s, B^s$ include \textit{smooth fit segments} $A_j^s, B_j^s$ where $j$ runs for total number of data segments.

  \item
	\textbf{Computing weighted correlation coefficient, $r^{w}$, for data segments:}
To compare the light curves we employ the weighted correlation coefficient (Equation~\ref{equ:sample_w_cross}). 
For the specific time lag $\Delta t$ we calculate the weighted correlation coefficient $r_{A^s, B_j}^{w} (\Delta t)$ between $j$th segment of light curve $B$ with those smooth fit segments of light curve $A^s$ which there is overlap between them.
To include the error bar of the data we choose the weight terms $w_i$ equal to inverse variance of data points i.e. $w_i=1/\sigma^2_i$ ($\sigma_i$ is the associated error bar of $i^{th}$ data point of $B_j$)\footnote{The normalized weight $w'_i$ can be derived through $w'_i=w_i / \sum\limits_{i=1}^n w_i$.}. 
Therefore for each data segment we can calculate $r_{A^s, B_j}^{w} (\Delta t)$ for different values of time lag $(\Delta t)$. Figure~\ref{fig:timedelay1} (plots 1) shows seven curves of $r_{A^s, B_j}^{w} (\Delta t)$ for seven segments of data $B_j$. Applying similar procedure on the data segments $A_j$ and smooth fit $B^s$ results in seven $r_{B^s, A_j}^{w} (\Delta t)$ curves illustrated in Figure~\ref{fig:timedelay1} (plots 2).

  \item
	\textbf{Computing weighted summation, $S^{p}$:}
We have mentioned the number of data points $n_d$ involved in correlation coefficient calculation determines the reliability of the correlation estimation. In other words the correlation coefficients corresponding to more number of data points in overlapped region is more reliable than those come from very few data points.
Therefore we incorporate the number of data points by using \textit{weighted summation} for combing the correlation coefficients of data segments.
The weighted summation of $r_{A^s, B_j}^{w} (\Delta t)$ correlation coefficients is given by
\begin{equation}\label{equ:w_sum1}
S_{A^s, B}^{p^k}(\Delta t)= \sum\limits_{j}  p^k_j (\Delta t)  r_{A^s, B_j}^{w} (\Delta t) 
\end{equation}
and similarly for $r_{B^s, A_j}^{w} (\Delta t)$ is
\begin{equation}\label{equ:w_sum2}
S_{B^s, A}^{p^k}(\Delta t)= \sum\limits_{j}  p^k_j (\Delta t)  r_{B^s, A_j}^{w} (\Delta t) 
\end{equation}
where in both equations $p^k_j (\Delta t)$ represent weights which can vary over time lag.
In this analysis we consider three different weights $p_j^{(1)}(\Delta t)=1$, $p_j^{(2)}(\Delta t)=({n_d}_j (\Delta t))^{\frac{1}{2}}$ and $p_j^{(3)}(\Delta t)={n_d}_j (\Delta t)$. Using these three weights and equations~\ref{equ:w_sum1} \& \ref{equ:w_sum2} we have two series of weighted summation i.e. $S_{A^s, B}^{p^{(1)}}(\Delta t), S_{A^s, B}^{p^{(2)}}(\Delta t), S_{A^s, B}^{p^{(3)}}(\Delta t)$ (Figure~\ref{fig:timedelay1}; plots 3, 6 and 9) and $S_{B^s, A}^{p^{(1)}}(\Delta t), S_{B^s, A}^{p^{(2)}}(\Delta t), S_{B^s, A}^{p^{(3)}}(\Delta t)$ (Figure~\ref{fig:timedelay1}; plots 4, 7 and 10).  
The maximum value of the weighted summation corresponds to the best time delay estimation. Hence we have six time delay estimations as $\tilde{\Delta t}_{A^{s}, B}^{p^{(1)}}, \tilde{\Delta t}_{A^{s}, B}^{p^{(2)}}, \tilde{\Delta t}_{A^{s}, B}^{p^{(3)}}$ and $\tilde{\Delta t}_{B^{s}, A}^{p^{(1)}}, \tilde{\Delta t}_{B^{s}, A}^{p^{(2)}}, \tilde{\Delta t}_{B^{s}, A}^{p^{(3)}}$ which are indicated in associated plots in Figure~\ref{fig:timedelay1}.

  \item
	\textbf{Mirror time delay estimator, $\tilde{\Delta t}^{\text{mirror}}_{A,B}$:}
Theoretically pairs of weighted summation $S_{A^s, B}^{p^{(k)}}(\Delta t), S_{B^s, A}^{p^{(k)}}(\Delta t)$ (for $k=1,2,3$) must show the similar trend in opposite directions with respect to time delay axes such that the associated best time lag values $\tilde{\Delta t}_{A^{s}, B}^{p^{(k)}}$ and $\tilde{\Delta t}_{B^{s}, A}^{p^{(k)}}$ should have close absolute values with opposite signs\footnote{Throughout this paper the sign of time delay estimations are chosen such that $\tilde{\Delta t}_{A, B}<0$ means light curve B is lagged with respect to light curve A.}.

This leads us to a new time delay estimator. We add $S_{A^s, B}^{p^{(k)}}(\Delta t), S_{B^s, A}^{p^{(k)}}(\Delta t)$ (for $k=1,2,3$ separately) curves considering the inverse time axis and divide the result by 2 to obtain an overall weighted summation versus time $S_{A, B}^{p^{(k)}}(\Delta t)$. 
The maximum value of the obtained overall weighted summation yields the \textit{mirror time delay estimation}, $\tilde{\Delta t}^{p^{(k)}\text{mirror}}_{A,B}$. Figure~\ref{fig:timedelay1}, plots 5, 8 and 11 illustrate results from mirror time delay estimator for $k=1,2,3$ respectively.
Among the three mirror time delay estimations we choose $\tilde{\Delta t}^{p^{(1)}\text{mirror}}_{A,B}$ (Figure~\ref{fig:timedelay1}; plot5) since it follows the form of the basic definition in statistics and also shows high performance results (Section~\ref{subsec:performance}). We use the rest of the results for reliability test of our time delay estimation as we will describe later in this Section.

  \item
	\textbf{Weighted average correlation, $\bar{S}^{p}$:}
For checking the reliability of estimated time delay we will need a normalized maximum weighted summation (between -1 to +1). This can ensure that the best estimated time delay corresponds to a reasonable correlation between the two light curves. Dividing the maximum weighted summation of correlations by sum of corresponding weights yields to \textit{weighted average correlations} in the form of 
\begin{equation}\label{equ:w_ave1}
\bar{S}_{A^s, B}^{p^k}(\tilde{\Delta t}_{A^{s}, B}^{p^{(k)}})= \frac{S_{A^s, B}^{p^k}(\tilde{\Delta t}_{A^{s}, B}^{p^{(k)}})}{\sum\limits_{j} p^k_j (\tilde{\Delta t}_{A^{s}, B}^{p^{(k)}})}
\end{equation}
similarly
\begin{equation}\label{equ:w_ave2}
\bar{S}_{B^s, A}^{p^k}(\tilde{\Delta t}_{B^{s}, A}^{p^{(k)}})= \frac{S_{B^s, A}^{p^k}(\tilde{\Delta t}_{B^{s}, A}^{p^{(k)}})}{\sum\limits_{j} p^k_j (\tilde{\Delta t}_{B^{s}, A}^{p^{(k)}})}
\end{equation}
and also for mirror estimator we have 
\begin{equation}\label{equ:w_ave3}
\bar{S}_{A, B}^{p^k}(\tilde{\Delta t}^{p^{(k)}\text{mirror}}_{A,B})= \frac{S_{A, B}^{p^k}(\tilde{\Delta t}^{p^{(k)}\text{mirror}}_{A,B})}{\sum\limits_{j} p^k_j (\tilde{\Delta t}^{p^{(k)}\text{mirror}}_{A,B})}
\end{equation}
where above Equations are valid for $k=1,2,3$ and weighted average correlation have the values in the range of $(-1, +1)$.

It is worth noting that using the weighted summation of correlations instead of weighted average correlation enables us to consider the number of segments of the data that contribute in estimation of the best time delay. This gives more weights to time delays that are derived using more segments of the data in comparison to time delays that are derived by considering only one or two segments of the data. Therefore we obtain the best time delay using maximum weighted summation while the weighted average correlation shows the reliability of estimation.
We see in our analysis that this approach can improve our analysis to minimize the possibility of having any outlier in our results.

  \item
	\textbf{Reliability test:}
To check the \textit{reliability of time delay estimation} we consider several criteria. First we follow \cite{aghamousa_timedelay_1} in assigning 
$\bar{S}_{A^s, B}^{p^k}(\tilde{\Delta t}_{A^{s}, B}^{p^{(k)}})$ \& 
$\bar{S}_{B^s, A}^{p^k}(\tilde{\Delta t}_{B^{s}, A}^{p^{(k)}})$ \& 
$\bar{S}_{A, B}^{p^k}(\tilde{\Delta t}^{p^{(k)}\text{mirror}}_{A,B})$
 $> 0.6$ ($k=1,2,3$) as an acceptance limit for correlation values. Next we check the consistency of each pair of $\tilde{\Delta t}_{A^{s}, B}^{p^{(k)}}$, $\tilde{\Delta t}_{B^{s}, A}^{p^{(k)}}$ time delays by checking their sign. We expect the consistent pairs owns opposite signs. In addition we calculate the mean $\overline{|\Delta t|}$ and standard deviations $\text{sd}(|\Delta t|)$ of absolute values of three $\tilde{\Delta t}_{A^{s}, B}^{p^{(k)}}$, $\tilde{\Delta t}_{B^{s}, A}^{p^{(k)}}$ time delay pairs and expect the reliable results show low discrepancy with $\text{sd}(|\Delta t|)<10$ days and $\text{sd}(|\Delta t|) / \overline{|\Delta t|} < 0.2$.
If any of the above conditions is not satisfied we consider our method disqualified to attain a reliable time delay estimation for the case of study. 

\end{itemize}

To sum up, our new algorithm offers the $\tilde{\Delta t}^{p^{(1)}\text{mirror}}_{A,B}$ as the time delay estimation provided passing reliability conditions successfully. 
In the next Section we discuss about Microlensing issue and the way our algorithm deals with that. We also evaluate the performance of the algorithm using TDC1 simulations in Section~\ref{subsec:performance} before deploying it on the lensed quasar SDSS J1001+5027 light curve data.

\subsection{Microlensing issue}\label{subsec:Microlensing}
Apart from discussed statistical issues, microlensing is another concern in time delay estimation which can cause distortion in the light curves and hence systematic errors in time delay estimation (e.g. \cite{Tewes2013a, Sluse2014}). In our analysis the light curves are compared in segments of data, considering that microlensing effect can be \textit{assumed} as the linear distortion in different segments. This is very important point in our approach since the correlation coefficient is unchanged under linear transformation of variables i.e. $m'=a \cdot m +b$ where $a$,$b$ are constant while $m$ and $m'$ are magnitudes of light curve before and after transformation receptively (Section~\ref{subsec:cross}). TDC1 blind analysis confirmed this point. With these in mind, we expect microlensing can not affect our time delay analysis in cases that pass all reliability tests and criteria (Section~\ref{subsec:time_delay}). Applying our algorithm on TDC1 data (Section~\ref{subsec:performance}) there are many cases that the results of the analysis cannot pass all the criteria and this could be (in cases) due to high frequency microlensing. But the main point here is that for those cases that could pass successfully our reliability conditions the results are in great agreement with true time delays which means the algorithm could take care of possible microlensing in these cases of data.

\subsection{Performance evaluation}\label{subsec:performance}
\begin{figure}
   \includegraphics[width=\textwidth]{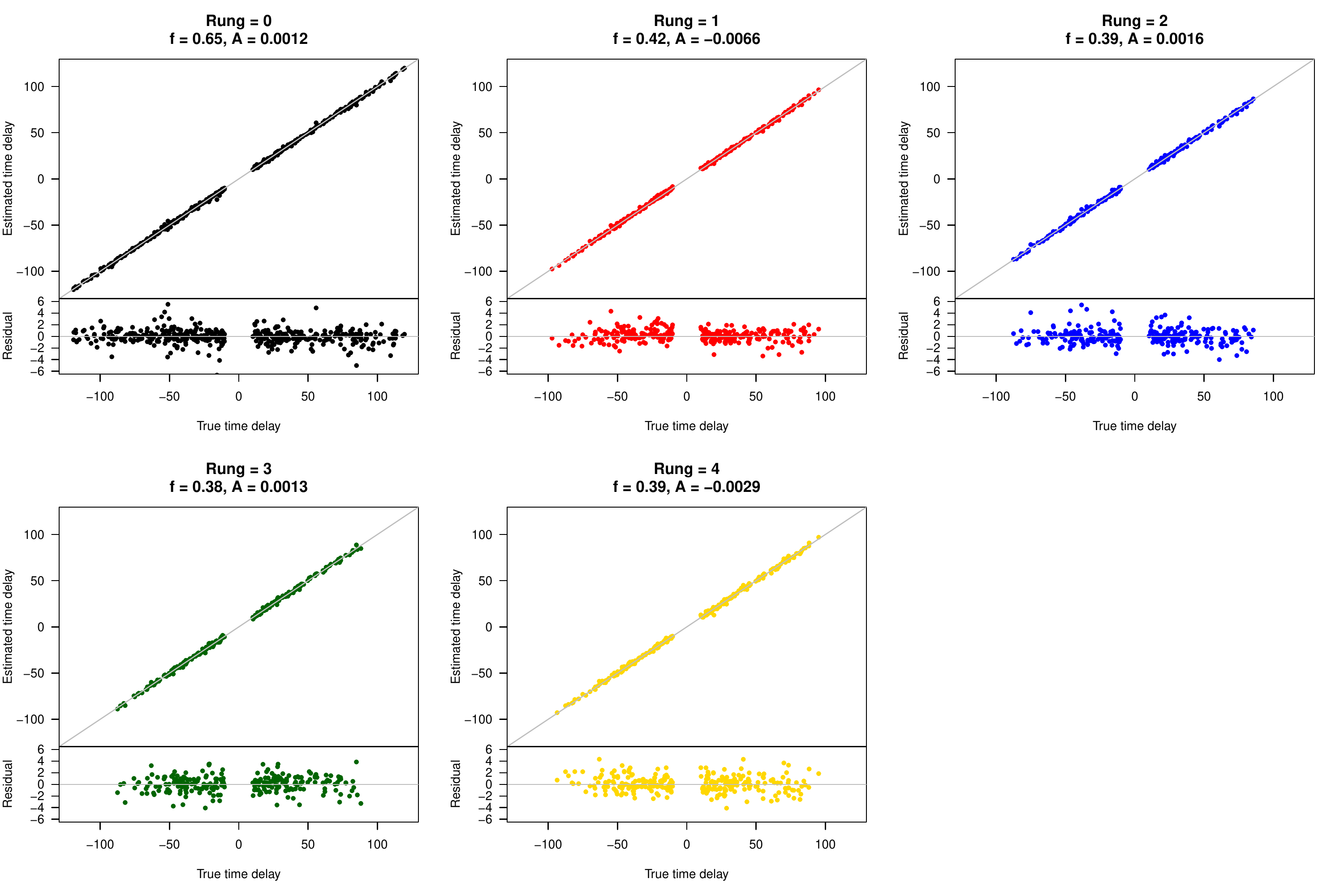}
  \caption{\label{fig:tdc1} The results of applying mirror estimator on TDC1 double system simulated data. 
In each plot the upper graph shows the estimated time delays versus true time delays while lower graph presents the residual of estimated and true time delays versus true time delays. The associated $f$ number and accuracy, $A$, are also reported in the titles. We see the time delay estimations in all rungs are in agreement with true time delays with no outlier.}
 \end{figure}

Our previous time delay estimation algorithm introduced in \cite{aghamousa_timedelay_1} has shown acceptable accuracy with no outlier. For comparison and check the performance of our new estimator, mirror estimator, explained in in Section~\ref{subsec:time_delay} we apply this procedure on TDC1 double systems simulated data. 
The TDC1 simulated light curves have been provided in five categories (called rung 0 to rung 4). As we see in Figure~\ref{fig:tdc1} the mirror estimator does not have outliers in results and all the time delay estimations are in agreement with true time delays. To quantify this evaluation we employ two metrics used in TDC1, the \textit{success fraction} or \textit{completeness},
\begin{equation}
f\equiv \frac{N_{submitted}}{N}
\end{equation}
and the \textit{accuracy} or \textit{bias},  
\begin{equation}
A=\frac{1}{fN}\sum_{i}\left ( \frac{ \tilde{\Delta t_i}-\Delta t_i}{\Delta t_i} \right )
\end{equation}
where $N$ is the total number of the light curves for analysis, $N_{submitted}$ is the number of submitted time delays, $\Delta t_i$ is true time delay and $\tilde{\Delta t_i}$ is estimated time delay.
Table~\ref{table1} shows these metrics measurements of the results of mirror time delay estimations in comparison with the results of our old algorithm \citep{aghamousa_timedelay_1} both applied on TDC1 double system simulated data. We can see dramatic improvements in new algorithm results. The accuracy increases around 10 times while completeness is also increased in all rungs in comparison to our old algorithm.

\begin{table}[!htb]
\centering
\begin{tabular}{l|l|l|l|l|}
\cline{2-5}
                             & \multicolumn{2}{l|}{mirror estimator}          & \multicolumn{2}{l|}{old algorithm}            \\ \cline{2-5} 
                             & \multicolumn{1}{c|}{f} & \multicolumn{1}{c|}{A} & \multicolumn{1}{c|}{f} & \multicolumn{1}{c|}{A} \\ \hline
\multicolumn{1}{|l|}{rung=0} & 0.65                   & 0.0012                 & 0.59                   & -0.0172                \\ \hline
\multicolumn{1}{|l|}{rung=1} & 0.42                   & -0.0066                & 0.39                   & -0.021                 \\ \hline
\multicolumn{1}{|l|}{rung=2} & 0.39                   & 0.0016                 & 0.38                   & -0.0239                \\ \hline
\multicolumn{1}{|l|}{rung=3} & 0.38                   & 0.0013                 & 0.36                   & -0.0206                \\ \hline
\multicolumn{1}{|l|}{rung=4} & 0.39                   & -0.0029                & 0.37                   & -0.0228                \\ \hline
\end{tabular}
\caption{Evaluation of mirror estimations: The measurements of metrics $f$ the success fraction and $A$ accuracy for the results of mirror estimations and our old algorithm \citep{aghamousa_timedelay_1} both applied on TDC1 double system simulated data. We see improvements in the results of mirror estimator in comparison to the old algorithm. The accuracy increases around 10 times in all the rungs of data along with an increasing in $f$ number values.}\label{table1}
\end{table}

\subsection{Error estimation}\label{subsec:error_estimation}
The error estimation denotes the assessment of uncertainty of an estimation. For a point estimation like our case of study this error would be reported with an interval around the estimate $\hat{\theta}\pm^{e_u}_{e_l}$, (where $e_l,e_u>0$). This interval which is associated with a probability $\psi \%$ is well known as \textit{confidence interval}. It implies if we repeat the experiment then associated interval $(a,b)$ captures the true value $\theta$ in $\psi \%$ of times\footnote{Note that in above definition $a$ and $b$ are random variables and can be changed between different experiments.}. It means
\begin{equation}
\label{eq:ci1}
\mathbb{P}(a\le \theta \le b) \equiv \psi \%
\end{equation}
where $a=\hat{\theta}-e_l$ and $b=\hat{\theta}+e_u$. Usually the confidence interval $(a,b)$ is constructed such that each wing around the $\hat{\theta}$ is equivalent to probability of $\frac{\psi}{2} \%$. 
The $68 \%$ confidence interval (1-$\sigma$ error bar) for an unbiased estimator with Normal distribution 
would have the simple form of
\begin{equation}
\label{eq:ci2}
\mathbb{P} \left( \hat{\theta} -\sigma \le \theta \le \hat{\theta}+\sigma \right) \equiv 68 \%
\end{equation}
where $\text{Var}(\hat{\theta})=\sigma^2$. 

However, in general to establish the $\psi \%$ confidence interval for a biased estimator with a non-Gaussian distribution we consider the expected value $\mathbb{E}(\hat{\theta})$ and an interval around it such that each side contributes $\frac{\psi}{2} \%$ of probability. Therefore corresponding estimation $\hat{\theta}$ occurs in this interval with $\psi \%$ chance by definition. It means
\begin{equation}
\label{eq:ci3}
\mathbb{P}\left( \mathbb{E}(\hat{\theta})-d_1 \le \quad \hat{\theta}  \quad \le \mathbb{E}(\hat{\theta})+d_2 \right) \equiv \psi \%
\end{equation}
where $d_1, d_2 >0$. By subtracting the true value and slightly manipulation we have
\begin{eqnarray}
\label{eq:ci4}
\mathbb{P} \left( (\mathbb{E}(\hat{\theta}_n)-\theta) -d_1 \le \hat{\theta}_n-\theta \le (\mathbb{E}(\hat{\theta}_n)-\theta)+d_2 \right)  &\equiv& \psi \% \nonumber \\
\mathbb{P} \left(\hat{\theta}-d_2-(\mathbb{E}(\hat{\theta})-\theta) \le \theta \le \hat{\theta}+d_1-(\mathbb{E}(\hat{\theta})-\theta) \right) &\equiv& \psi \%
\end{eqnarray}
The last equation introduces the $\psi \%$ confidence interval $(a,b)$, where $a=\hat{\theta}-d_2-\text{Bias}(\hat{\theta})$, $b=\hat{\theta}+d_1-\text{Bias}(\hat{\theta})$ and $\text{Bias}(\hat{\theta})=\mathbb{E}(\hat{\theta})-\theta$. Equivalently it presents the error of estimation in the form of $\hat{\theta}\pm^{e_u}_{e_l}$ where $e_l=d_2+\text{Bias}(\hat{\theta})$ and $e_u=d_1-\text{Bias}(\hat{\theta})$.\footnote{A similar issue has been discussed in \cite{Wasserman2004} (chapter 8).}

We can see the error depends on bias term $\text{Bias}(\hat{\theta})=\mathbb{E}(\hat{\theta})-\theta$ and $d_1, d_2$ which are determined by \textit{unknown} distribution of estimator. In our time delay analysis through the simulations we are able to obtain an \textit{approximate} distribution of the time delay estimators and their associated bias values. This leads to an approximate error estimation for the derived time delay. The most critical point in the error estimation based on simulations is using more realistic simulated data to obtain more accurate distribution of estimator and hence error estimation. It is also worth noting that although a symmetric distribution of estimator leads to equal $d_1$ and $d_2$ but the bias of estimator results an asymmetric error bar ($e_l \neq e_u$) for estimation. We will see the effect of these facts in the results of our analysis in this work.

\subsection{Simulations}\label{subsec:simulations}
To generate a pair of simulated light curve data for a doubly strong lens system we first employ a \textit{simple} Monte Carlo simulation procedure.
We assume a smooth fit as the true light curve and then add the Gaussian noise according to error bar of the data to obtain the simulated light curve $A_{sim}$. To obtain the second simulated light curve $B_{sim}$, we move the smooth light curve in time axes equal to assumed time delay and 
repeat the previous step to generate corresponding simulated data. In Section~\ref{sec:data} we mentioned the SDSS J1001+5027 light curves suffer from gaps in the data. This causes the large uncertainty in the smooth fit between the data segment regions such that the corresponding continuous smooth fit cannot be trusted to be used in the simulation. 
Therefore we need a well sampled light curve data with sufficiently small time intervals to use its corresponding smooth fit as the true light curve in simulation. To overcome this issue we employ one of the synthetic strong lens light curve data from Time Delay Challenge (TDC0) \citep{TDC0-2013}.  The simulated light curves in one of the categories of data (rung 0) in this challenge are generated in equispaced time epoch with one day interval. 
We select the light curves labeled by pair 4 in this category which is similar to SDSS J1001+5027 light curves data and take associated smooth fit as the true fit in simulation. We generate 5000 pairs of simulated light curves data $A_{sim}, B_{sim}$, with different random true time delays in the range of $-120$ days to $-100$ days, using the actual error bars of SDSS J1001+5027 light curves at same time epochs. We should emphasize here that all simulated data in our analysis would have the characteristics of the real SDSS J1001+5027 data (we used the same time spacing of data epochs and error characteristics) and we used TDC0 sample just to come up with a realistic true quasar light curve. 

\begin{figure}
   \includegraphics[width=\textwidth]{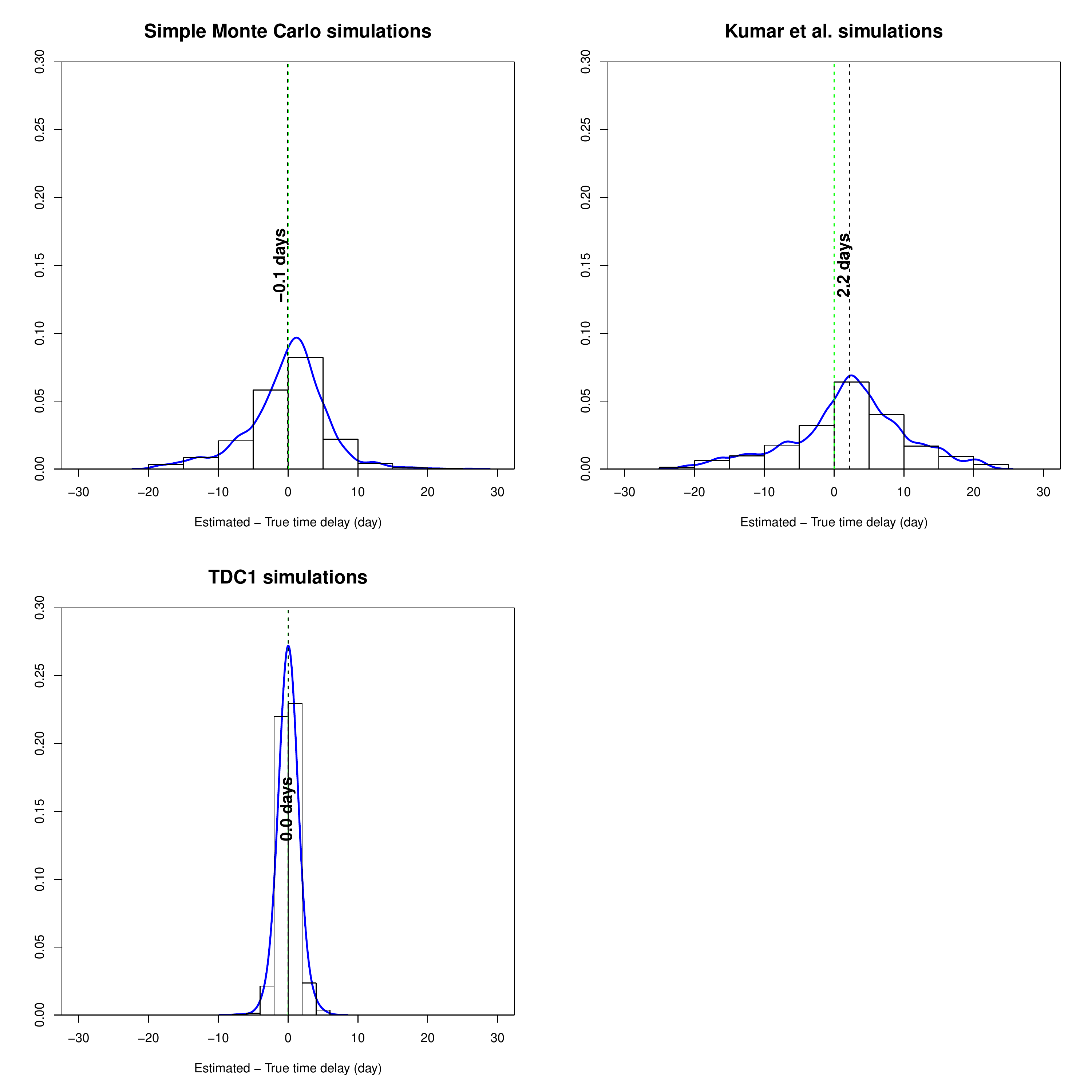}
  \caption{\label{fig:distributions} The kernel density estimation and histogram of difference between estimated and true time delays for mirror estimator using simple Monte Carlo simulations (left-upper panel) and simulations by \cite{COSMOGRAIL_XIV} (right-upper panel) and TDC1 double systems simulations (lower panel). The first two sets of simulations generated by considering the characteristics of lensed quasar SDSS J1001+5027 light curves while the TDC1 simulations reflect the general properties of double lensed quasar light curves from near future observations.
The black dashed lines represent bias values, expected values of difference between estimated and true time delay, while the zero value is indicated by green dashed line in each plot.}
 \end{figure}

We apply the mirror time delay estimator on simulated light curves data in the same procedure explained in Section~\ref{subsec:time_delay}. 
Figure~\ref{fig:distributions} (left-upper panel) illustrates the distribution of difference of mirror estimations and corresponding true time delays. The lower panel also shows similar distribution comes from TDC1 double systems simulated data discussed in Section~\ref{subsec:performance}.

For comprehensiveness we have also employed the simulations provided by \cite{COSMOGRAIL_XIV} for the SDSS J1001+5027 system since they have claimed considering various possible systematics in the data (and hence in their simulations). These simulated light curves are generated with different assumed true time delays from $-135.6$ days to $-105.2$ days. We apply the mirror estimator on these simulations and obtain distribution of time delay estimations subtracting corresponding true time delays (Figure~\ref{fig:distributions}; right-upper panel).

The difference in distributions reflects the different procedure used to generate the sets of simulations.
In Section~\ref{sec:results} we use these distributions to estimate the error of time delay estimation of SDSS J1001+5027 lensed system.

\section{Results and Discussion}\label{sec:results}
\begin{figure}
   \includegraphics[width=\textwidth]{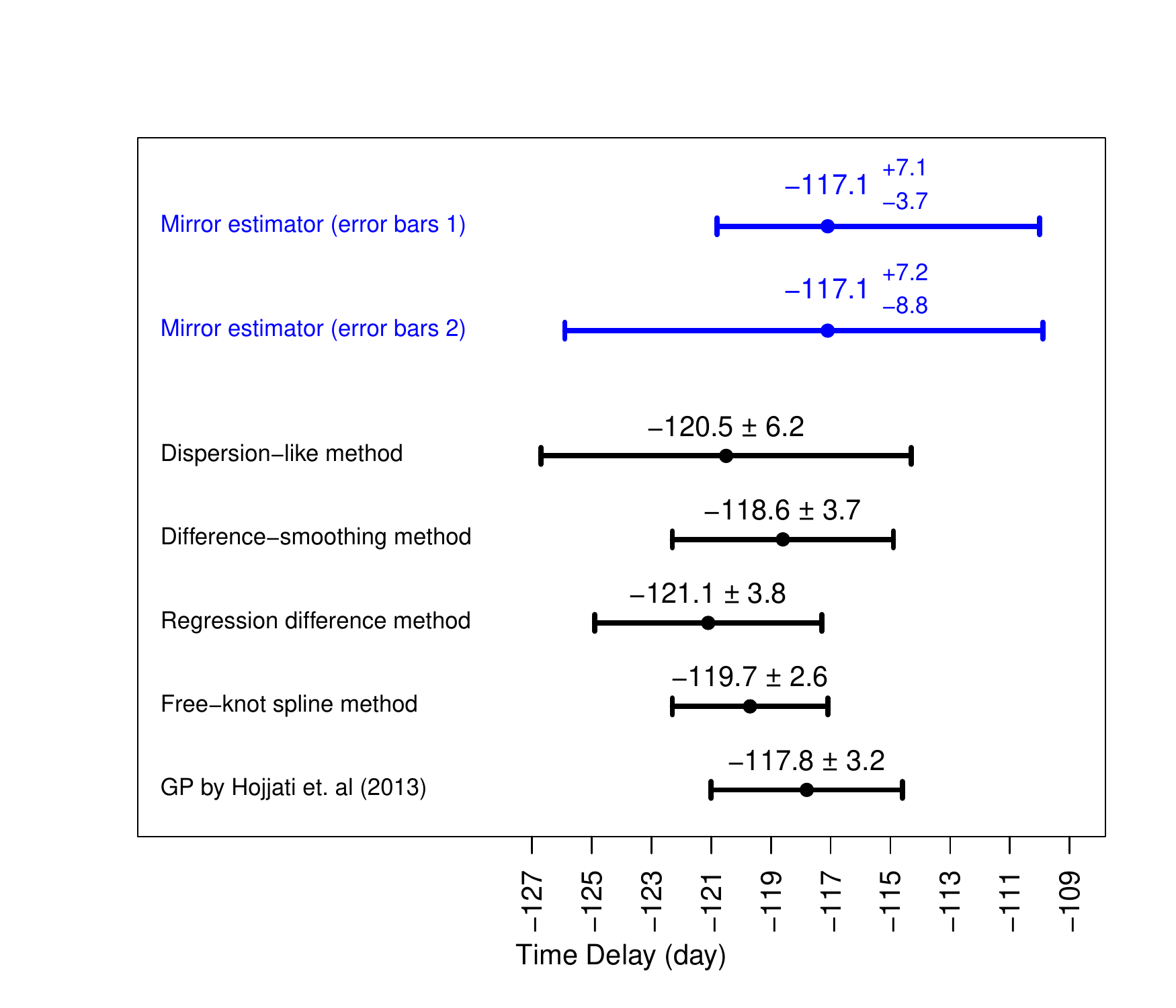}
  \caption{\label{fig:timedelay} The mirror time delay estimation of lensed quasar SDSS J1001+5027 with two different error bars using different simulations in comparison with five other estimations reported by \cite{COSMOGRAIL_XIV}. The error bars using simple Monte Carlo simulations and simulated light curves generated by \cite{COSMOGRAIL_XIV} are titled by error bars 1 and 2 respectively in the graph.}
 \end{figure}

Applying mirror estimator on lensed quasar SDSS J1001+5027 light curve data with passing all reliability criteria (Section~\ref{subsec:time_delay}) successfully we obtain the time delay equals to -117.1 days. Following error estimation procedure explained in Section~\ref{subsec:error_estimation} by using distributions of estimator from different sets of simulations (Figure~\ref{fig:distributions}; upper panels) result in $\tilde{\Delta t}^{\text{Monte Carlo}}_{A,B}=-117.1^{+7.1}_{-3.7}$ days and $\tilde{\Delta t}^{\text{Kumar}}_{A,B}=-117.1^{+7.2}_{-8.8}$ days. 
In addition using distribution of TDC1 simulations (Figure~\ref{fig:distributions}; lower panel) leads to uncertainty less than 1.5 days which shows the possibility of more precise estimation providing high quality data. 
As we see using different simulated light curves yields different size of error bars specifically simulations by \cite{COSMOGRAIL_XIV} result the largest error bars. This somehow reflects the fact that simulation of the light curves of strong lens systems can be difficult and one has to consider many details and unknowns. Figure~\ref{fig:timedelay} illustrates the mirror time delay estimation with two different error bars in comparison with five other time delay estimations reported by \cite{COSMOGRAIL_XIV}. We can see the time delay estimations using different algorithms shows approximately a range between $-120$ and $-114$ days which are in agreement with our estimation. These results show the need of increasing the quality of the data to have more accurate results.  

\section{Conclusion}\label{sec:conclusion}
In this paper we present an algorithm for time delay analysis of strong lensed systems applying it on doubly lensed quasar SDSS J1001+5027.
This algorithm is a modification of our previous formalism introduced in \cite{aghamousa_timedelay_1} which is mainly based on iterative smoothing and cross-correlation. 
Using smoothing method we obtain the smooth light curves $A^s, B^s$ of corresponding $A, B$ light curves. We compare then the light curves in segments by calculating weighted cross-correlation between $A$ and $B^s$ and also between $B$ and $A^s$ for different time delays. As the first improvement in this analysis we utilized the weighted correlation for including the error bar of the data epochs.
Finally we have two sets of correlation coefficients from comparing segments of the data and corresponding smooth fits for each time delay.
The number of data points in overlapped region in two segments of data has important effect in reliability of the obtained correlation coefficient. Therefore we consider the number of data points in adding correlation coefficient of segments of the data by using weighted summation such that the weight terms are a function of number of data points. Using this procedure we average correlation values from the data segments to provide two similar trends of correlation coefficient values versus time delay in opposite directions. Having these, we propose a new time delay estimator, namely mirror estimator, where we consider an important list of criteria to check the reliability of the estimation. 

For evaluating the performance of the proposed modified algorithm, we apply it on all TDC1 double system simulated data which we already used in our previous analysis. The results show impressively $\sim10$ times improvement in accuracy and also increasing in completeness of time delay estimations in comparison with the results of our old algorithm. For estimating the uncertainties of estimation we need to obtain the distribution of estimations. For this purpose we use simple Monte Carlo simulation based on assumed true light curves and other simulations provided by \cite{COSMOGRAIL_XIV} both generated in order to resemble lensed quasar SDSS J1001+5027. In addition we use distribution of estimations coming from TDC1 double systems simulated data. 

Applying mirror estimator on SDSS J1001+5027 data lead to $\tilde{\Delta t}^{\text{Monte Carlo}}_{A,B}=-117.1^{+7.1}_{-3.7}$ days and $\tilde{\Delta t}^{\text{Kumar}}_{A,B}=-117.1^{+7.2}_{-8.8}$ days using two different sets of simulations for error estimation. 
Furthermore using TDC1 simulations which present the general properties of double lensed quasar light curves we expect from near future observations results to the uncertainty of less than 1.5 days.
Using different simulations for error estimation we obtain different size of error bars. We can somehow conclude from all these that the simulated data provided by \cite{COSMOGRAIL_XIV} might be a more conservative representation of the actual data in comparison with simple Monte Carlo simulations. 
The large uncertainties in our estimated time delay for this case of study is also somehow reflected in the different time delay estimations using other techniques by different groups. We expect that increasing the quality of the light curves will definitely results in more precise time delay estimations that can be essential for application in cosmology.

We truly appreciate the comments and suggestions by our unknown referee that enhanced the quality of our work.
We thank \cite{COSMOGRAIL_XIV} authors for sharing simulated light curves of lensed quasar SDSS J1001+5027.
We thank Malte Tewes, Eric Linder, Vivien Bonvin, S. Rathna Kumar and Dominique Sluse for various discussions and their comments.
A.A. acknowledges support from the Korea Ministry of Science, ICT and Future Planing at the Asia Pacific Center for Theoretical Physics.
A.S. would like to acknowledge the support of the National Research Foundation of Korea (NRF-2016R1C1B2016478).

\bibliographystyle{apj}
\bibliography{main}

\end{document}